\documentstyle[preprint,prl,aps,epsf]{revtex}
\begin{document} 
\draft

\title{Low-Temperature Phase Transition in
Bi$\rm_2$Sr$\rm_2$Ca(Ni$\rm_x$Cu$\rm_{1-x}$)$\rm_2$0$\rm_8$: Evidence for
Unconventional Superconductivity}

\author{R. Movshovich,$^1$ M. A. Hubbard,$^2$ M. B. Salamon,$^2$ A. V.
Balatsky,$^1$ 
R. Yoshizaki,$^3$ and J.~L.~Sarrao,$^1$}
\address{$^1$Los Alamos National Laboratory, Los Alamos, NM 87545 \\
$^2$University of Illinois at Urbana-Champaign, Urbana, Illinois 61801-3080
\\ $^3$Institute of Applied Physics,
Cryogenics Center, University of Tsukuba, Tsukuba, Japan}

\date{September 4, 1997}
\maketitle

\begin{abstract} 

We report the discovery of a low-temperature phase transition in
Bi$\rm_2$Sr$\rm_2$Ca(Ni$\rm_x$Cu$\rm_{1-x}$)$\rm_2$0$\rm_8$ high temperature
superconductor. This transition manifests itself as a sharp reduction of the
thermal conductivity of the samples at a temperature $\rm T_c^\ast \approx
200\ mK$. The temperature dependence of the thermal conductivity changes
dramatically from $T^\alpha$ with $\alpha$ between 1.6 and 1.75 above the
transition to T-linear behavior below the transition. Application of a small
magnetic field suppresses the low-temperature phase. We interpret this
behavior as a phase transition into a second bulk low-temperature
superconducting state in which the time-reversal symmetry is likely to be
broken. This discovery constitutes direct evidence for unconventional
superconductivity in high temperature superconductors. 
\end{abstract}


\narrowtext
 
Unconventional superconductivity is a very active area of research, both in
cuprate and heavy-fermion compounds. These two classes of materials hold the
possibility that the attractive interaction between charge carriers
originates not from the electron-phonon interaction, as in conventional
superconductors, but rather from the magnetic fluctuations present in these
systems. Such an interaction may result in a superconducting state with the
symmetry of the order parameter lower than that of the underlying lattice.
These states are called unconventional, as opposed to the s-wave
phonon-mediated conventional superconductors. If a particular compound has
more than one superconducting phase, at least one of these phases must be
unconventional, because there is only one possible symmetry for the s-wave
state: that of the underlying crystallographic lattice. Therefore, the
presence of multiple superconducting phases is a sufficient but not a
necessary condition for an unconventional state. Until recently, only one of
the suspected unconventional superconductors -
UPt$\rm_3$\cite{stewart84:UPt3} - has been unambiguously shown to have more
than one superconducting phase. It was first shown that there are two phase
transitions out of the zero-field low-temperature superconducting state when
magnetic field is increased~\cite{qian87,muller87,schenstrom89}; then two
transitions were resolved in zero-field heat capacity
measurements~\cite{fisher89}; and finally a complete phase diagram was
constructed using ultrasound velocity~\cite{shireen90}.  These results were
immediately accepted as proof that superconductivity in UPt$\rm_3$ is
unconventional (for a recent review of the heavy-fermion superconductivity,
see Ref. \cite{heffner96}). An unambiguous observation of multiple
superconducting states in high temperature superconductors would settle the
question of whether superconductivity in this class of compounds is  indeed
unconventional.

In this article we report on the observation of a second low-temperature
superconducting phase transition in Ni-doped
Bi$\rm_2$Sr$\rm_2$CaCu$\rm_2$0$\rm_8$ high temperature superconductor. The
low-temperature phase transition manifests itself as a dramatic drop of
thermal conductivity between $T_c^\ast = 200$ mK and 160 mK. Our discovery
of multiple superconducting phases in a high temperature superconductor
unambiguously answers the question of whether superconductivity in this
class of compounds is of unconventional character: it is. There is by now a
large body of experimental work on high temperature superconductors that was
interpreted as evidence of unconventional d-wave pairing. It includes
observation of a phase shift of $\pi$ in the corner dc SQUID
experiments~\cite{wallman93}, linear-in-temperature microwave penetration
depth in the low temperature limit~\cite{hardy93}, and large gap anisotropy
via angle-resolved photoemission~\cite{shen93}. Our results give support to
the conclusions about the unconventional character of the high temperature
superconductors that were drawn from these measurements. 

The Bi-2212 samples used for thermal conductivity measurements were prepared
by the traveling solvent floating zone technique~\cite{yoshizaki}.
Individual samples were obtained by cutting parts of the rods out and
cleaving them down to an appropriate size. The samples used for thermal
conductivity measurements were thin rectangular slabs, with thickness
varying between 10 $\mu$m and 80 $\mu$m and a surface area of  about 1 mm
$\times$ 5 mm. The nominal concentrations of Ni dopant substituted for Cu
(which was determined by the ratios of the starting materials) varied
between 0\% and 2.4\% for the two series of samples investigated. We
characterized the samples by measuring resistance and magnetization.
Superconducting transition temperatures were obtained as the intersection
points of linear fits to the magnetization curve in the region of its rapid
change below the transition and the flat background above the transition.
All of the samples used for thermal conductivity measurements had magnetic
transition widths of about 1 K, which indicates the high quality and
homogeneity of the samples. During the growth of the sample by the traveling
solvent floating zone technique, impurities may be redistributed along the
sample rod. For example, impurities may be pushed preferentially to the ends
of the sample rod. The samples used in this work had the following nominal
Ni concentrations and T$\rm_c$'s: 0\% Ni - T$\rm_c$ = 89 K, 0.6\% Ni -
T$\rm_c$ = 78 K, 1.5\% Ni -  T$\rm_c$ = 77 K, and 2.4\% Ni - T$\rm_c$ = 74
K.  The superconducting transition temperatures reflect the true
concentration of the Ni impurities better then the nominal Ni
concentrations. Therefore we will identify different samples by their
T$\rm_c$'s in the rest of the article. 

We used a standard steady state ``two heaters, one thermometer" method to
collect the thermal conductivity data. One end of the sample is thermally
sunk to the bath,  and the thermometer is thermally attached to the opposite
end. Two resistive heaters are thermally attached at different points along
the sample between the bath and thermometer contacts. By separately turning
the heaters on and measuring thermometer's temperature in both cases one can
easily deduce the thermal conductivity of a section of a sample between the
two points where the heaters are thermally attached. Silver pads for thermal
contacts to bath, heaters and thermometer were evaporated onto the samples
and annealed at 300$^\circ$C for two hours. We used silver paint to attach
samples to the bath. Annealed 25 $\mu$m-diameter platinum wires provided
thermal links from the sample to the thermometer and to the two heaters. 

\begin{figure}
\epsfxsize=6in
\epsfbox{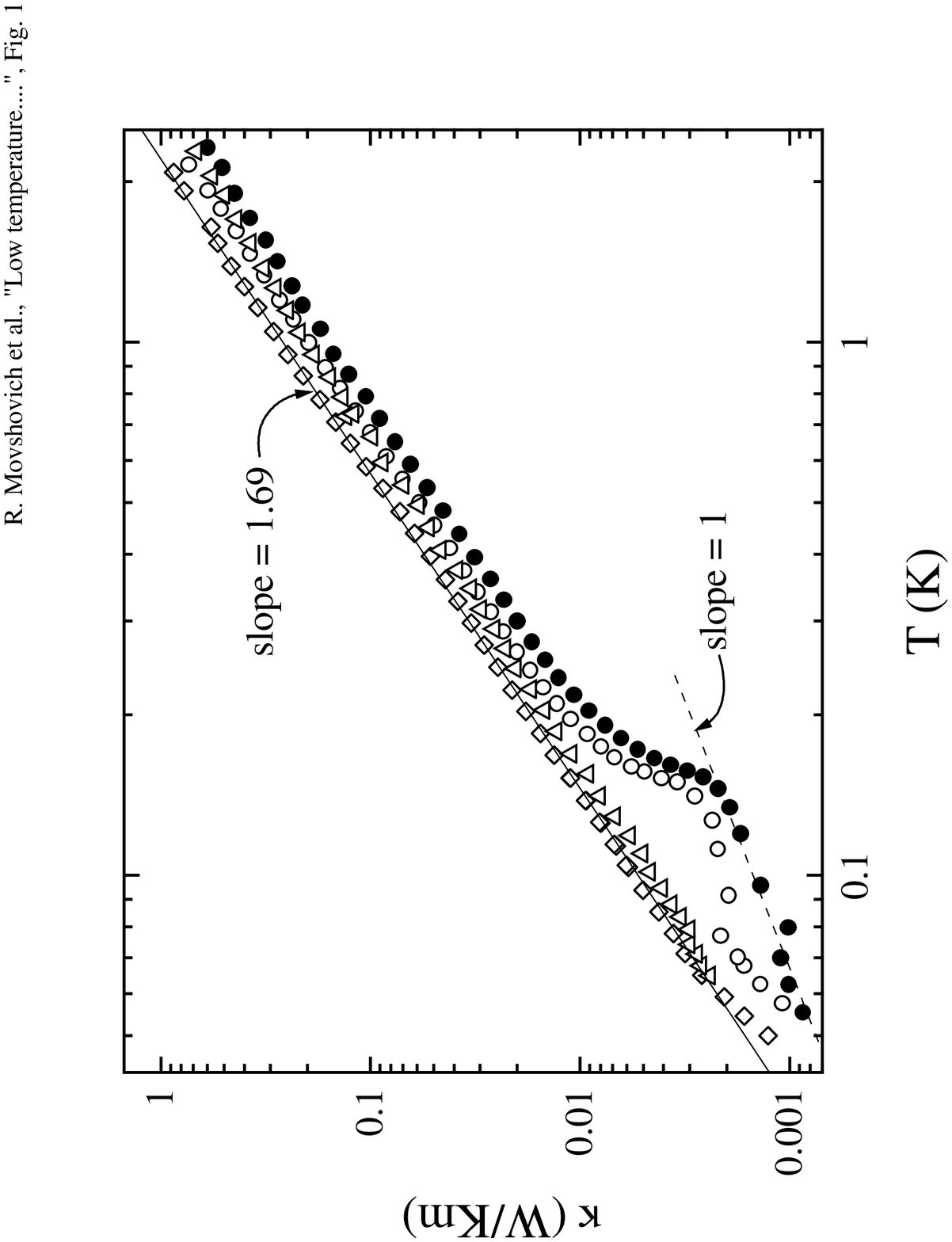}
\caption{Thermal conductivity of several Ni-doped
Bi$\rm_2$Sr$\rm_2$CaCu$\rm_2$0$\rm_8$ samples. 
$\Diamond$ - undoped, $T\rm_c = 89 K$; $\circ$ - $T\rm_c = 78K$; $\bullet$ -
$T\rm_c = 77K$; $\triangle$ - $T\rm_c = 74K$. }
\label{thermal conductivity - all}
\end{figure}

Fig.~\ref{thermal conductivity - all} shows the thermal conductivity data as
a function of temperature on a log-log plot for several samples with
different nominal Ni concentration. The samples in Fig.~\ref{thermal
conductivity - all} are identified by their upper superconducting transition
temperatures, which reflect the true level of impurity concentration. The
two samples that display phase transitions have T$\rm_c$ = 77 K and 78 K. We
observed the T$_c^\ast$ = 200 mK  transition in the  sample with T$\rm_c$ =
77 K on three different cooldowns, in two different thermal conductivity
cells, with two different sets of  platinum thermal leads, and two different
thermal connections between the sample and the bath. The low-temperature
phase transition therefore is a robust and reproducible property of the sample.

The large size of the drop in thermal conductivity indicates that the
transition takes place within the main thermal transport channel of the
system.  Thermal conductivity $\kappa$ is a sum of two components: $\kappa =
\kappa_{el} + \kappa_{ph}$, where $\kappa_{el}$ and $\kappa_{ph}$ are
electronic and phonon thermal conductivity. We can make an upper limit
estimate the low-temperature phonon thermal conductivity in Bi-2212 by using
the expression $\kappa_{ph} = {1 \over 3}\beta \langle {\it v}_{ph}
\rangle \Lambda_0 T^3$\cite{thacher67}, similar to the calculation of this
limit for YBCO~\cite{taillefer97}. Here $\beta$ is a coefficient of the
$T^3$ (phonon) term of the heat capacity, $\langle {\it v}_{ph} \rangle
={\it v}_L(2s^2+1)/(2s^3+1)$, with $s = {\it v}_L/{\it v}_T$,
the ratio of the longitudinal to transverse velocities\cite{thacher67}. For
single crystals $\Lambda_0 = 2\overline{\it w}/\sqrt{\pi}$, where
$\overline{\it w}$ is the geometric mean width of a rectangular sample.
Using the tabulated values for longitudinal and transverse sound velocity as
well Debye temperature $\Theta_D$~\cite{dominec93} ($\Theta_D$ is used to
calculate $\beta$), we estimate the phonon contribution to the thermal
conductivity to be 7 times smaller in Bi-2212 than in YBCO for samples with
the same physical dimensions. Specifically for the 1.5\% Ni-doped Bi-2212
sample (T$\rm_c$ = 77 K) that shows a second low-temperature phase
transition, we calculate the phonon thermal conductivity to be $\kappa_{ph}
= 0.2 \times T^3$ W/mK$\rm^4$, or 1.6 mW/Km at T = 200 mK. Comparing this
number with the experimental value of $\kappa$ = 9 mW/Km  at T = 200 mK, we
see that phonon contribution is a small fraction of the total thermal
conductivity and cannot account for the drop of about 7 mW/Km between 200 mK
and 150 mK. Therefore, the phase transition at T$\rm_c^\ast$ = 200 mK must
have an electronic origin. 

A slowly opening energy gap that accompanies a second-order superconducting
phase transition leads to a smooth decrease of the electronic thermal
conductivity as a function of temperature. This is the behavior exhibited by
most of the conventional superconductors like Al, Sn, Nb, In, and others.
Therefore, the smooth change in thermal conductivity in Fig. 1 is indicative
of the second-order character of the low-temperature phase transition. An
alternative explanation of rounding of the thermal conductivity data at the
phase transition is an inhomogeneous distribution of the Ni impurities. Such
an explanation appears unlikely in view of the extremely sharp character of
the high temperature superconducting transitions in both of the samples
displaying the second low-temperature phase transition. 

The data in the high-temperature phase (both for the samples with and
without the low-temperature phase) between 50 mK and 2 K can be described
very well  by a single power law $\kappa = T^\alpha$ with $\alpha$ between
1.6 and 1.75. However, as Fig. 2 shows, the low temperature data can be
equally well described as the sum of a linear and quadratic temperature
dependence. Indeed, the latter term follows directly from the theory of
electronic thermal conductivity of Bardeen, Rickayzen, and
Tewordt\cite{bardeen59} applied for a pure 2D d-wave superconductor in the
limit $T \ll \Delta$~\cite{krishana97}. The $T^2$-dependence is due to the
low energy quasiparticle states in the nodes of the $d_{x^2-y^2}$ order
parameter. 

Below the low-temperature transition the nearly $T^2$-term is quickly
suppressed and the thermal conductivity data become linear in temperature.
This feature is emphasized in the main body of Fig.~\ref{therm_cond in
field}, which shows thermal conductivity divided by temperature as a
function of temperature. In a pure conventional s-wave superconductor with a
fully gapped Fermi surface all of the quasiparticle states are frozen out
far below the superconducting transition temperature T$_c$, and all of the
thermal transport is due to phonons, with the phonon thermal conductivity
$\kappa_{ph} \propto T^3$. As was discussed in the context of conventional
low-temperature superconductors, introduction of pair-breaking magnetic
impurities leads to the formation of intra-gap impurity states\cite{YSR}
(Shiba states) and impurity bands (for sufficiently large impurity
concentration). The impurity band can lead to the formation of tails in the
DOS that extend to zero energy \cite{YSR,Balatsky:preprint97}, leading to a
nonzero DOS at E = 0. If the scattering is strong, the impurity band is
close in energy to the mid-gap point and produces a roughly
energy-independent DOS. Such a constant DOS in the region about E = 0 makes
an impurity band metallic in character at low temperature, with the
metallic-like thermal conductivity $\kappa$/T = constant. This is exactly
what we observe  experimentally, and therefore the thermal conductivity data
in the low temperature state is consistent with that of a superconducting
state with a fully gapped Fermi surface and an intra-gap impurity band.
Opening of the gap eliminates the low-energy nodal quasiparticle states of
the $d_{x^2-y^2}$ order parameter that lead to $T^2$ contribution to the
thermal conductivity above T$_c^\ast$. The change in the temperature
dependence from nearly quadratic above T$\rm_c^\ast$ to linear below it
reflects the qualitative change of the spectrum of quasiparticle states as
the system undergoes the phase transition.

\begin{figure}
\epsfxsize=6in
\epsfbox{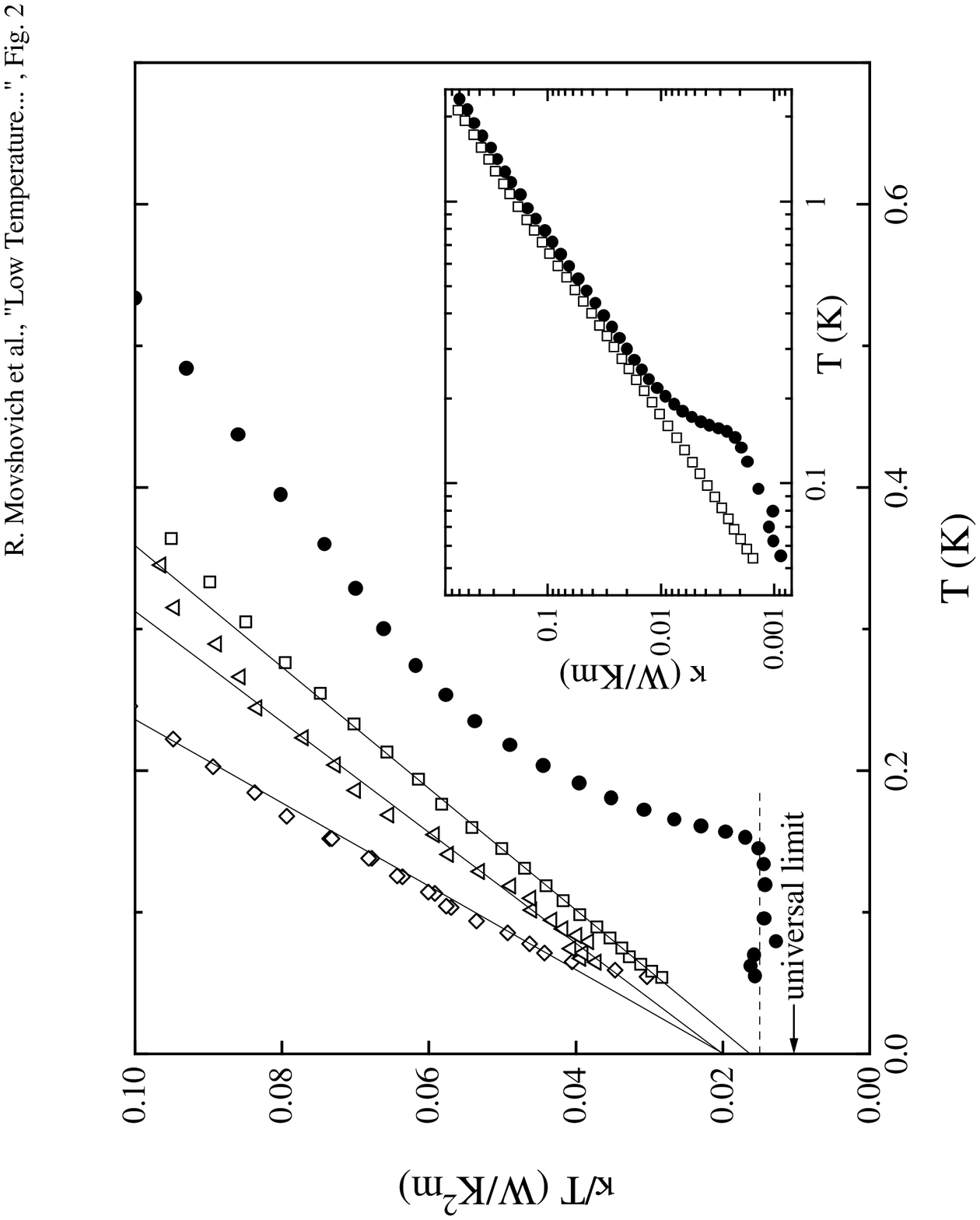}
\caption{Thermal conductivity divided by temperature as a function of
temperature. $\Diamond$ - undoped, $T\rm_c = 89 K$; $\triangle$ - $T\rm_c =
74K$; $\bullet$ - $T\rm_c = 77K$, $H = 0 G$; $\Box$ - $T\rm_c = 77K$, $H
\approx 200 - 300 G$. Inset: thermal conductivity of
Bi$\rm_2$Sr$\rm_2$CaCu$\rm_2$0$\rm_8$ sample with T$\rm_c = 77 K$. $\circ$ -
$\rm H = 0 G$, $\Box$ - $H \approx 200 - 300 G$. }
\label{therm_cond in field}
\end{figure}

To investigate the nature of the low-temperature transition, we repeated
thermal conductivity measurements on a sample with T$\rm_c$ = 77 K with a
magnetic field of several hundred Gauss applied by placing a small permanent
magnet near the sample. The results of this measurement are shown in
Fig.~\ref{therm_cond in field}. The magnetic field suppressed the phase
transition. The values extrapolated to $T = 0$ are close to the universal
low temperature thermal conductivity limit for doped
Bi$\rm_2$Sr$\rm_2$CaCu$\rm_2$O$\rm_8$~\cite{herschfeld86,schmitt86,graf96:tc,taillefer97,movshovich97:2} in a d-wave phase, which is also indicated in
Fig~\ref{therm_cond in field}. However, since the density of states in an
impurity band depends on the impurity concentration, we do not expect the
thermal conductivity in the low-temperature phase with fully gapped Fermi
surface to be universal in the low temperature limit.

The inset of Fig.~\ref{therm_cond in field} shows a log-log plot of the
thermal conductivity as a function of temperature for one of the samples
that underwent the low-temperature phase transition. The data in field
follow a single power law between 50 mK and 2 K as for the samples that did
not display the low-temperature phase transition. Detailed quantitative
investigation of the effect of the magnetic field is in progress.

All of the features of the data presented in our paper are consistent with a
second low-temperature superconducting transition. To summarize: 1) the
sharp drop in thermal conductivity indicates the loss of a substantial
number of primary heat carriers in the system, i.e. normal quasiparticles;
2) the different temperature dependence above and below the transition
implies qualitatively different quasiparticle excitation spectra in the two
phases, indicating different order parameters; 3) the linear-in-temperature
dependence of the thermal conductivity in the low-temperature phase is
consistent with the metallic-like contribution of the impurity band in a
fully gapped superconductor; 4) a modest magnetic field appears to suppress
the low-temperature phase. Once it is established that the second phase is
in fact a different superconducting phase, the rest of the argument follows:
superconductivity is unconventional because there can not be two different
superconducting states with the same symmetry of the underlying
crystallographic lattice in the same compound. Therein lies the significance
of our results: one can make a general statement about (un)conventionality
of a superconducting state based on the discovery of two different
superconducting phases in a compound. The statement is that the
superconductivity in such a compound is indeed unconventional. Such a
conclusion does not depend on the exact microscopic nature of the first or
second superconducting state.

There are several possible scenarios that may be responsible for the
low-temperature phase transition. One of them is the s-wave pairing of the
quasiparticles in the nodes of the d-wave gap (which does not have to be
phonon-mediated). Another is a new phase that is stabilized by interaction
of the condensate with magnetic field produced by Ni impurities. The
presence of pairbreaking impurities  (Ni is most likely a magnetic impurity
in Bi-2212~\cite{ishida93,kitaoka93}) leads to the finite DOS for nodal
quasiparticles which in turn can participate in the second phase transition
at $\rm T_c^\ast \approx 200\ mK$. However, magnetic impurities also act as
pair breakers for s-wave superconducting states, whereas nonmagnetic
impurities do not~\cite{anderson59}. Therefore, within the s-wave scenario,
increasing the concentration of Ni impurities leads to a competition between
an increased DOS that are able to take advantage of pairing and increased
pairbreaking and  quasiparticle scattering rate. This competition may
explain the absence of the low-temperature phase both in the undoped and in
the 2.4\%Ni-doped samples (see Fig.~\ref{thermal conductivity - all}): the
undoped sample does not have a large enough low-energy DOS, and the
pairbreaking effect of impurities dominates in the 2.4\% Ni-doped sample,
suppressing the second phase.  

When the d-wave nodes are gapped at the low-temperature phase transition,
which is indicated by the qualitative change in the temperature dependence
of the thermal conductivity data, it is likely that the low-temperature
state breaks time reversal symmetry. For example, consider going from a
d-wave state to a sum of a d-wave and an s-wave order parameters. The nodes
of the $d_{x^2-y^2} + s$ order parameter are simply moved to slightly
different positions on the Fermi surface (for $|s| \ll |d|$) from their
positions for the $d_{x^2-y^2}$ superconducting state. Therefore, to
completely gap the nodes of the $d_{x^2-y^2}$ order parameter, the s-wave
component must be added with a different phase (as an imaginary part), i.e.
the low-temperature (complex) order parameter must have a $d + is$ symmetry
to allow for a non-zero gap value everywhere on the Fermi surface. The same
is true for an order parameter that is the sum of two different d-wave
components, $d_{x^2-y^2}$ and $d_{xy}$. Such a state must have a $d + id$
symmetry to have a non-zero gap value everywhere on the Fermi surface.  Such
a complex order parameter leads to broken time reversal symmetry, an
exciting possibility that should be reflected in a number of unusual
physical properties. Spontaneous orbital currents in such states were
considered originally by Anderson and Morel\cite{anderson61}. More recently
the signatures of broken time reversal symmetry were investigated in heavy
fermions\cite{sigrist91,sauls94}, where models that describe most of the
experimental observations in superconducting UPt$\rm_3$ lead to spontaneous
orbital moments of the Cooper pairs. 

Thermal conductivity is a bulk probe. We therefore conclude that the phase
transition we observe is an intrinsic bulk property of the Ni-doped Bi-2212.
For that reason, this low-temperature transition is different from the
recently reported anomalies in tunneling into
YBa$\rm_2$Cu$\rm_3$O$\rm_7$~\cite{covington:preprint97_1}, which can be
understood in terms of the appearance of the subdominant order parameter at
the surface of the high temperature superconductor. Surface-induced states
that violate time-reversal symmetry were discussed originally by Sigrist
{\it et al.}~\cite{sigrist95}. Generation of the secondary order parameter
due to Andreev bound states on the surface has been recently discussed by
Sauls and co-workers\cite{fogelstrom:preprint97}, who also considered broken
time-reversal symmetry for the case of such surface-induced states in d-wave
superconductors.

Recently, K. Krishana {\it et al.} reported an observation of an anomaly in
the thermal conductivity data as a function of magnetic field in single
crystals of pure Bi$\rm_2$Sr$\rm_2$CaCu$\rm_2$0$\rm_8$\cite{krishana97}.
They observe a sharp change in the slope  of the thermal conductivity with
magnetic field (field sweeps were performed at constant temperature ranging
from 6 K to 20 K). Thermal conductivity first decreases by a few percent
with increasing magnetic field up to a critical field H$\rm_c$ of several
Tesla, and then becomes field-independent. This anomaly may be a consequence
of the transition into a $d + i d$ state~\cite{laughlin97}. If the
low-temperature state in Ni-doped Bi$\rm_2$Sr$\rm_2$CaCu$\rm_2$0$\rm_8$ is
driven by the interaction between the magnetic field produced by the
Ni-impurities and the condensate, both of these phenomena might be related.

The low-temperature phase transition at T$_c^\ast \approx 200$ mK must be
accompanied by a feature in specific heat. We performed specific heat
measurements on 1\%Ni- and 2\%Ni-doped Bi-2212 samples grown by a self-flux
technique~\cite{kendziora}, which produced shiny, micaceous samples with a
typical surface area of 10 mm$\rm^2$. We used a semi-adiabatic heat pulse
method to collect the specific heat data.  We do not see any feature in the
data around 200 mK that would indicate a phase transition. A smooth
background results from a nuclear Schottky feature and a
linear-in-temperature term, similar to the data for pure
Bi-2212~\cite{baak93}. At 200 mK $C \approx 20 \rm\ mJ/mol\ K$, which is
about two orders of magnitude greater than specific heat values at this
temperature for typical metals. Given our null results, the size of the
feature in specific heat at the low-temperature transition is less then 0.2
mJ/molK (precision of our measurement is $\approx 1$\%). This makes
observation of the low-temperature phase transition via specific heat
measurement a very challenging experiment, and also highlights the utility
of thermal conductivity measurements as a probe of this transition. 

In conclusion, we have discovered a low-temperature phase transition at $\rm
T_c^\ast = 200$ mK in the high temperature superconductor
Bi$_2$Sr$_2$Ca(Ni$\rm_x$Cu$\rm_{1-x}$)$_2$0$_8$. The phase transition
manifests itself as a sharp drop in the thermal conductivity between $\rm
T_c^\ast \approx 200$ mK and 160 mK. This phase is suppressed by the
application of a weak magnetic field. We interpret our observations as
evidence for a second low-temperature superconducting phase in
Bi$_2$Sr$_2$Ca(Ni$\rm_x$Cu$\rm_{1-x}$)$_2$0$_8$. Such observation of
multiple superconducting phases in high temperature superconducting
compounds is a manifestation of the unconventional character of
superconductivity in these materials. This low-temperature superconducting
state probably breaks time-reversal symmetry.

Acknowledgements: We thank R. Ecke for his help with the data acquisition
hardware and for his invaluable assistance with the manuscript. We thank H.
Safar for his help with sample preparation. We are grateful to J. D.
Thompson and J. L. Smith for useful discussions. We thank J. A. Sauls, L.
Taillefer, and N. P. Ong for making available the preprints of their
articles prior to publication. D.~Hristova helped with the operation of the
cryogenic apparatus. Work at Los Alamos was performed under the auspices of
the U.S. Department of Energy. M. A. Hubbard and M. B. Salamon acknowledge
support by the National Science Foundation (DMR 91-20000) through the
Science and Technology Center for Superconductivity.


\begin{thebibliography}{10}
\bibitem{movshovich97:2}
detailed comparison of the data Ni-doped Bi-2212 with theoretical predictions
  for universal low-temperature conductivity will be presented elsewhere.
\end{thebibliography}

\end{document}